# Learners' Quanta based Design of a Learning Management System

S. Sengupta

Bengal Institute of Technology
Kolkata, INDIA

mesouvik@hotmail.com

N. Chaki
Dept. of CSE
University of Calcutta
Kolkata, INDIA

nabenduc@yahoo.com

R. Dasgupta

National Institute of Technical
Teachers' Training & Research
Kolkata, INDIA
ranjandasgupta@ieee.org

***Abstract:*** In this paper IEEE Learning Technology System Architecture (LTSA) for LMS software has been analyzed. It has been observed that LTSA is too abstract to be adapted in a uniform way by LMS developers. A Learners' Quanta based logical level design that satisfies the IEEE LTSA standard has been proposed for future development of efficient LMS software. A hybrid model of learning fitting into LTSA model has also been proposed while designing.

***Key Words:*** IEEE LTSA, e-Learning , LMS, SCROM, Hybrid Model

## I. Introduction

The non-linear way of storage of information in the form of hypertext has brought a revolutionary change in the teaching-learning process [13]. In the hypertext document, links have been established in such a way that the user can explore, browse and search for not only a particular item but can also get information regarding relevant/associated issues. Presentation of information through multiple media formats enriches users' experience and improves the learning process [20]. Cockertion and Shimell evaluated hypermedia document as a learning tool [18]. They have focused their study on hypermedia document and included graphical controls for simple interaction behavior. Vassileva and Deters designed a dynamic courseware generator tool based on AI planning techniques [14]. These works emphasize more on pedagogical and other issues rather than identification of requirements for a overall Learning Management System (LMS).

Keith S. Taber and his associates [15] at the University of Cambridge put forward a project aimed to integrate English and Science standards using technology as a vehicle. The emphasis, however, was merely to improve the presentation of the learning material. John Munro [17] of the University of Melbourne has also worked on identification of the requirements for the effective delivery of course content. He essentially tried to analyze learning based on some pre-defined key issues. Guanon Zhang [16] has designed a computer based knowledge system for assisting persons in making decisions and predictions upon human or data-mining knowledge. The identification of specific requirements for individual learners of vastly different background and the design of effective LMS is being considered as a challenging problem all over the globe. Several efforts in this regard have been reported which are mostly course-specific.

While working on design of Learners' Quanta (LQ) based efficient and adaptive learning system, it has been observed that a generic platform is required to design and develop an LMS with the adaptive algorithm already proposed in [4], [9] and [12]. A brief report of LQ based requirement specification is presented in section V for the sake of completeness. The working group IEEE "1484 Learning Technology Standards Committee (LTSC)" has designed an architecture called Learning Technology System Architecture (LTSA) to standardize web-based content delivery for all learning technology systems [3]. A high level design of the LTSA is now been proposed which can accommodate the LQ based learning system.

The five-layered LTSA standard specifies a high level architecture for information technology-supported learning, education, and training systems. This standard is pedagogically neutral, content-neutral, culturally neutral, and platform-neutral. The layer's specification is from abstraction towards implementation as one moves from Layer I to Layer V. The layer I is the highest level of abstraction where two entities Learner and Environment and their interactions are shown. Layer II is less abstract than Layer I and describes a generic five-step algorithm of learning technology considering the human-centered features. In layer III the system components like Learner, Evaluation, Knowledge Library, Assessment, and Learning Content are described. Layer IV specifies perspectives of different stakeholders. The final layer describes operational components like client-servers, interface protocols, name space URL etc.

Now this architecture cannot be considered as a blue print for designing a single system, but can be considered as a framework for designing a range of systems over time. Neither this framework specifies any implementation technologies. In this paper, a design which can be treated as *an implementable form* of LTSA has been proposed.

## II. Adaptation of LTSA by Popular LMS Software

The LMS software available in the markets are lacking from a common standardization of their architecture [1]. These software follow SCROM as the standard for their content management [6, 7, 8] but for the overall architecture, LTSA being so generic that it is very hard for different vendors to follow it in a universal way. Each of the market available LMS software has some merit and limitations. Major features of a few such software have been presented in Sec. II.A and Sec. II.B deals with their limitations. In Sec. III proposed modification of LTSA has been presented while Sec. IV deals with the UML design and its short description Sec. V as mentioned earlier depicts a brief idea about the LQ based requirement specification.

### A. Market available LMS Software

Moodle [6] is one of the common names in the LMS market. In this software, course listing shows descriptions for every course on the server, including accessibility to guests. Courses can be categorized and searched - one Moodle site can support thousands of courses. A teacher has full control over all settings for a course, including restricting other teachers. Choice of course formats such as by week, by topic or a discussion-focused social format. Assignments can be specified with due dates and maximum grades. Students can upload their assignments (any file format) to the server - they are date-stamped. Teacher feedback is appended to the assignment page for each student, and notification is mailed out. The teacher can choose to allow resubmission of assignments after grading (for re grading).

Sakai [7] is a set of software tools designed to help instructors, researchers and students to create websites on the web. For courseware, Sakai provides features to supplement and enhance teaching and learning. For collaboration, Sakai has tools to help organize communication and collaborative work on campus and around the world. Using a web browser, users choose from Sakai's tools to create a site that meets their needs, no knowledge of HTML is necessary. The powerful tools provided with Sakai are Syllabus tool, Schedule tool, Resources tool, Assignment tool, grade book tool, email archive tool, announcement tool etc [7].

Ilias [8] is a powerful Open Source Learning Management System for developing and realizing web-based e-learning. ILIAS offers multiple ways to deliver learning content. All types of document files can be uploaded; SCORM 2004, SCORM 1.2 and AICC are supported. ILIAS includes an internal authoring environment to create XML-based learning modules, that can include images, flash, applets and other web media files.ILIAS provides a "Personal Desktop" for every user with facilities like listing of selected courses, groups and learning resources, bookmark management, personal notes, external web feeds, calendar, personal learning progress and also has other important modules like course management, assessment evaluation, learning content or authoring for the use of the teachers or administrator.

### B. Limitations of the market available software

These software are not designed as per LTSA architecture but LTSA is so generic that one can easily adapt it to the architecture of the software to understand it better. All of the software have implemented the 'Layer I' well in their system. In 'Layer II' negotiate in learning style is not that well organized in these software. Though Moodle, Sakai provides the scope of interaction with the teacher while learning, but how to slow down the pace of learning or how to explain some key words better are not well designed in these systems. In 'layer III' the system components like Learning Content, Knowledge Library, Students' performance database are well identified in most of the software, but the approach of reusability has not been defined in any of the software. The 'layer IV' of LTSA architecture is well implemented in the leading software, at least three stake holders perspective student, teacher and administrator are taken care of all. Finally in 'layer V' Operational components are identified by: i) Systems, e.g., clients and servers ii) Connectors (interface protocols), e.g., HTTP, PPP iii) Busses (namespaces), e.g., URLs, telephone numbers. As all these LMS softwares are based on web architecture, they have implemented many of these operational components.

### III. LIMITATIONS OF LTSA AND THE PROPOSED DESIGN

Some of the functional areas not included in LTSA are identified and a brief report of the same is presented here:

a) The model does not regard the learning object designer as an integrated component in the learning process [2].
b) The students evaluation records are stored but how to use it is not specified.
c) For a distance mode learner, if the learner possess some fundamental wrong/incomplete idea and the feedback system fails to identify it, then the LTSA layer II algorithm falls under a never ending iterative cycle.
d) Students counseling is not included in the LTSA architecture. Students take on courses generally by only knowing the name of the course. Many times there are some prerequisites that they overlooked.

In the design we have introduced the hybrid mode of learning (E Learning + contact Learning) inside LTSA. If any learner fails to achieve the expected level of knowledge in one or multiple attempts in distance mode, there is a provision of contact classes for that learner on that subject to remove all the problems and confusions. Hybrid courses have the following inherent advantages over face-to-face teaching or totally online courses:

*Convenience:* Coursework accommodates students' schedules, plus commuting time is decreased. Hybrid instructors report increased interaction and contact among students and between the instructor and the students [10].

*Flexibility:* Instructors can accomplish certain learning objectives more successfully than in traditional courses because of the flexibility of the Hybrid model [11]**.**

*Increased learning:* Faculty almost universally reports that their students learn more in the Hybrid format than they do in traditional class sections. Instructors report that students write better papers, perform better on exams, produce higher quality projects, and are capable of more meaningful discussions on course material.

*Increased retention:* Data from the University of Central Florida (UCF) also show that student retention in Hybrid courses is better than retention in totally online courses and equivalent to that of face-to-face courses.

Jun [21] has presented an effect on the pair programming in offline or online environment for commercial programming. In our proposed design the students evaluation records are used for the evaluation of the excellence of the course, excellence of the Instructor (for contact classes) and counseling of that student in any other course. The learner counseling is also proposed which may be conducted by taking intelligent small test on their earlier knowledge or by navigating them through the list of prerequisites for different courses. The two-mode evaluation process (online and offline) is also introduced in our design - online objective test and offline assignments. The purpose of the evaluation system is not only to assist users in verifying their knowledge online but also to create a stimulating environment, where users improve their knowledge gradually[19].

## IV. ARCHITECTURAL DESIGN OF THE PROPOSED MODEL USING UML:

An architectural design of a Learning Management System using UML diagrams has been presented in this section. This design consists of one use case, one sequence and one activity diagram. The basic architecture behind the design is LTSA but augmented with some added features like hybrid model of learning, two mode evaluation process, counseling etc.

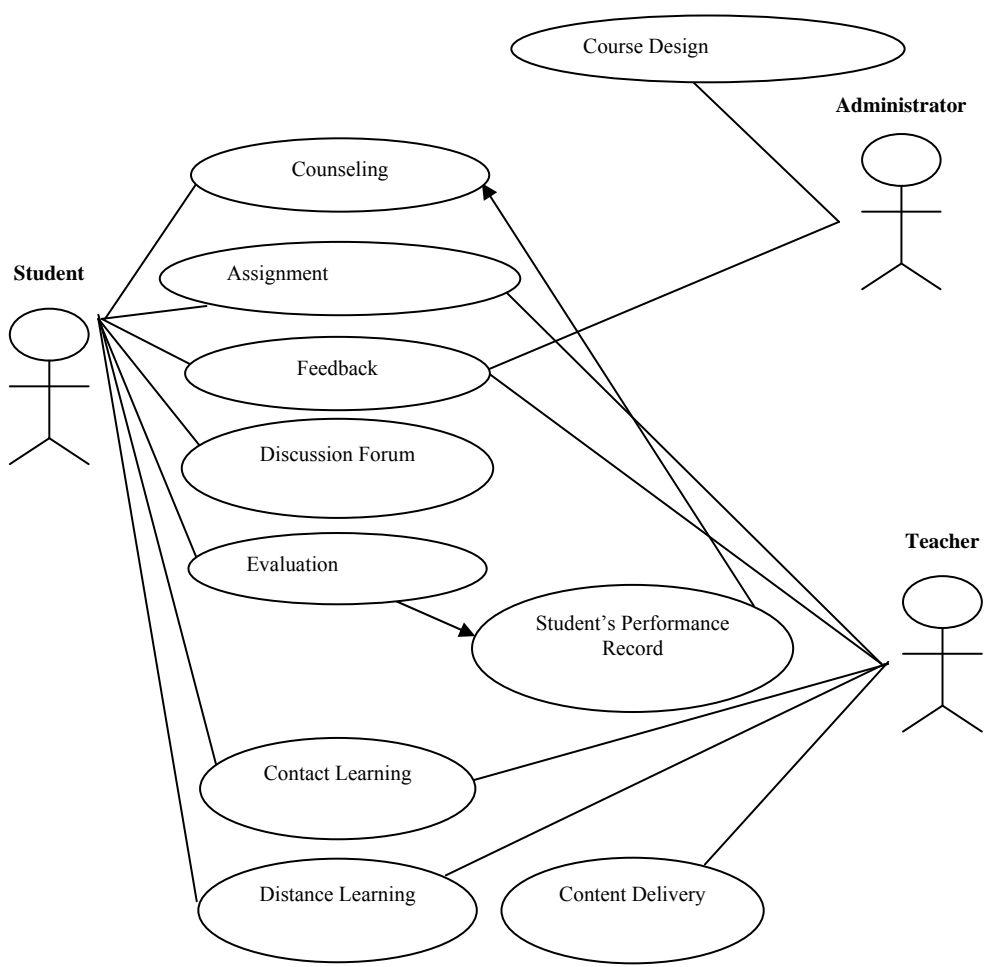

Fig 1: Use Case Diagram of LMS

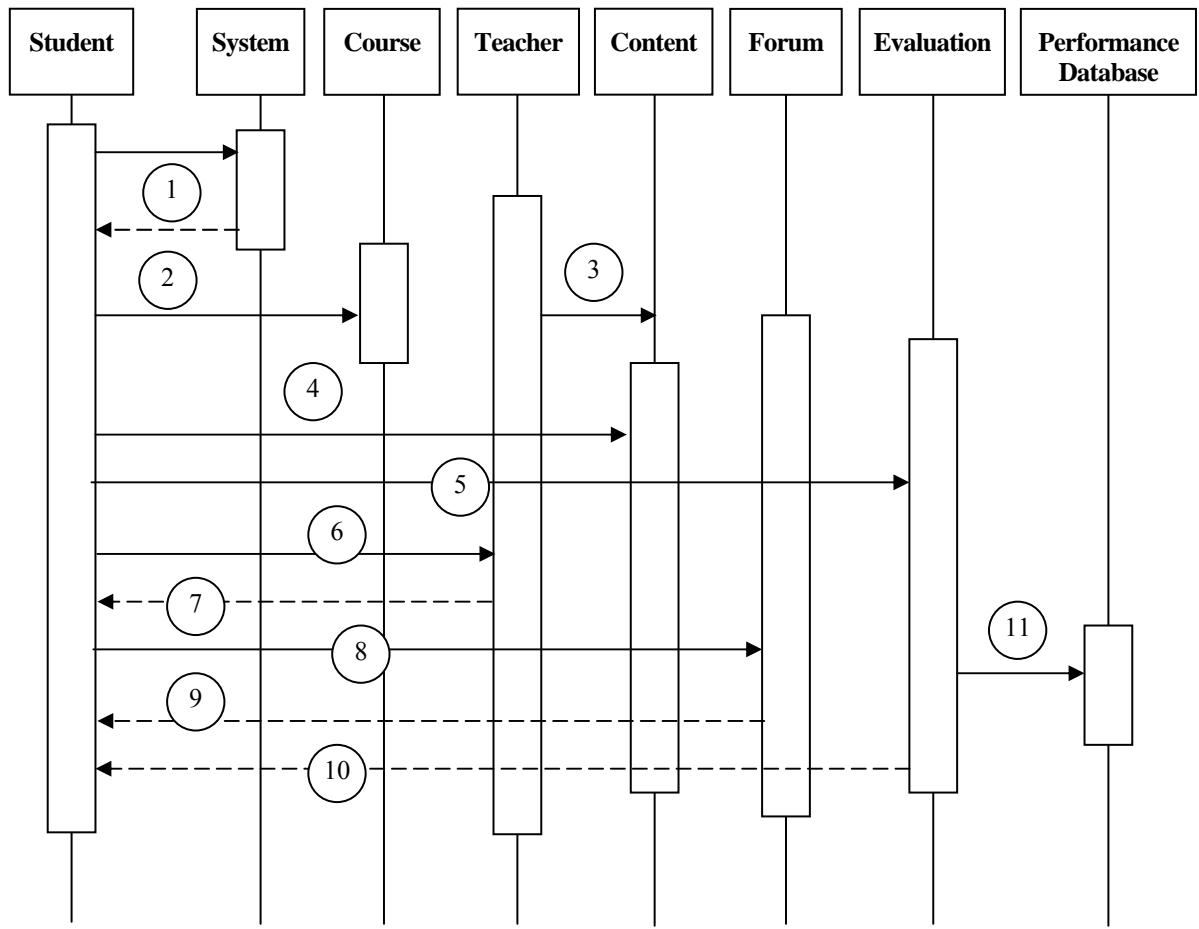

| 1 | Counseling |
| 2 | Enrollment |
| 3 | Delivery |
| 4 | Learning |
| 5 | Participate |
| 6 | Asking help |
| 7 | Negotiating Teaching Style |
| 8 | Contributory |
| 9 | Beneficiary |
| 10 | Result |
| 11 | Record Evaluation |

Fig 2: Sequence Diagram of LMS

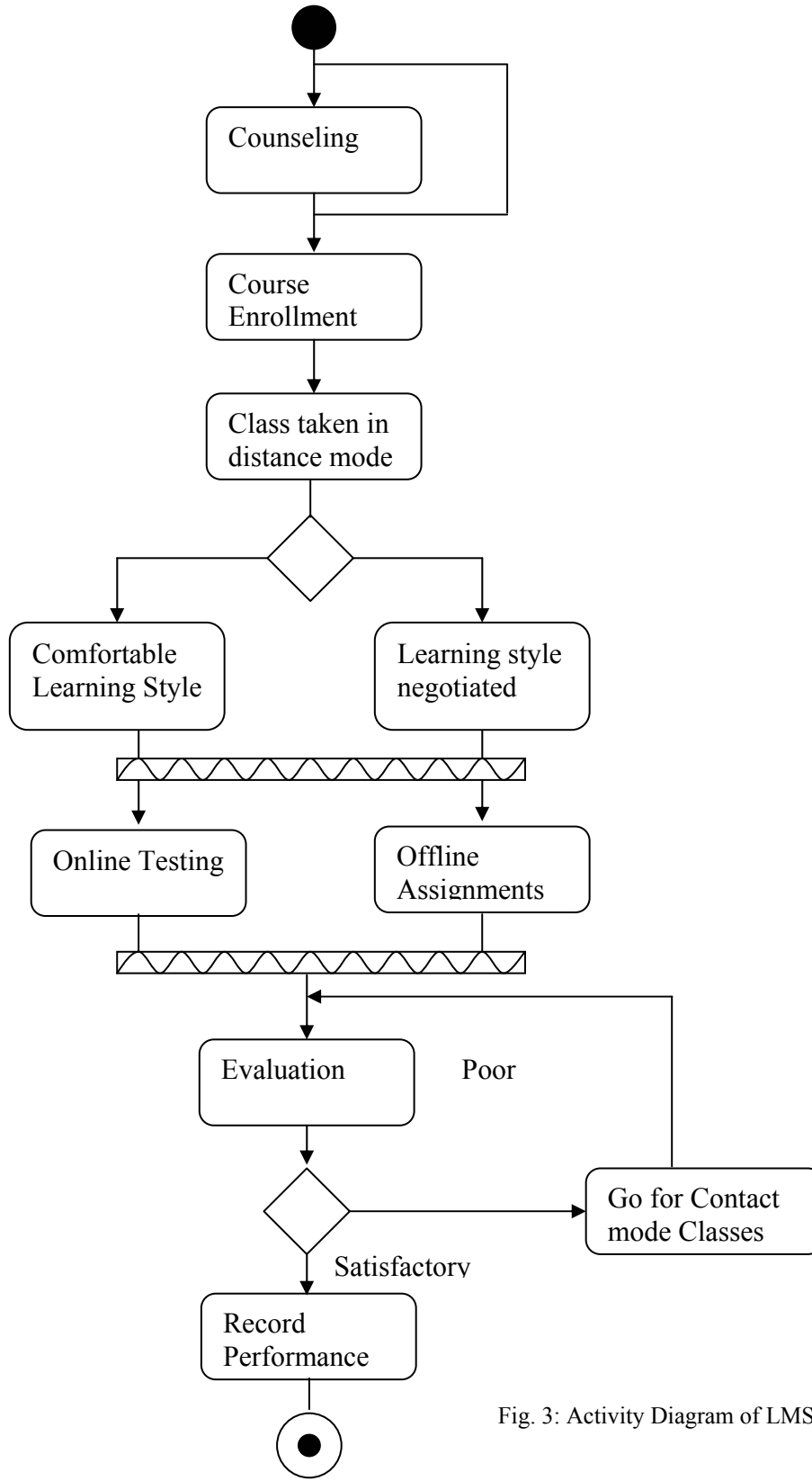

Fig. 3: Activity Diagram of LMS

In the use case diagram [fig 1] teacher, student and administrator are identified as the main actors and it supports both contact learning and distance learning (i.e. hybrid model). The students performance record is stored and used for counseling of the students for different courses or can be used for counseling different students on that course. Discussion forum is the place where a stakeholder like learner can change his role and can teach the other learners as per LTSA Layer IV specification.

In the sequence diagram [fig 2] the link between teacher and student shows that the student can negotiate in learning style in the mid of the course with the teacher, thus satisfying the part of layer II learning cycle.

In the activity diagram [fig 3] the counseling is the first activity even before the enrolment. This design specifies the distance learning to be the next step. However the learner can negotiate with learning style while learning. The performance of the learner evaluated in two ways. One by online test, mainly objective or MCQs and are more frequent, and other by offline test mainly submission of assignments, less frequently conducted. Next if the performance is not achieved up to a desired level then contact classes are arranged for that student on that course. Thus in this design the basic learning cycle described in LTSA layer II is combined with hybrid model of learning and two mode evaluation technique.

## V. LQ BASED REQUIREMENTS SPECIFICATION AND DESIGNING

### A. Terminology used

*Learner's Quanta (LQ):* A Learner's Quantum of study is a measured part of a topic with a specific output objective requiring a specific input knowledge on part of the learner [4, 9, 12].

*Learner's Quanta Cloud (LQC):* A Learners' Quanta Cloud (LQC) is a collection of semantically related group of learners' quanta. Any arbitrary quanta, $LQ_i$ could be part of more than one LQC, where LQs are grouped based on different semantics.

*Knowledge Factor (KF):* A Knowledge Factor (KF) is an atomic element of information. Each LQ is associated with a unique set of input KF and another set of output KFs. The intersection of the input and output set for a particular KF is usually a Null set.

*Target Knowledge Factors (TKF):* The Target Knowledge Factors are the set of KFs specified by the user as the set of output objectives he/she wishes to acquire.

*Known Knowledge Factors (KKF):* The Known Knowledge Factors are the set of KFs specified by the user as the set of already known elements of information for a particular learner.

*LQ Dictionary:* The LQ dictionary for a subject area refers to the entire set of LQs with their corresponding KFs stored in one specific place. The dictionary is different for each subject area.

Given

I:    Input knowledge set of the learner
       I is the set of KKFs.
R:    Requirement knowledge set specified by the learner
       R is the set of TKFs
Oi:   Objective set to be attained by a learner on completion of
       the ith LQ
Pi:   Pre-requisite knowledge set for the ith LQ

In a real life situation, the intersection of R and I would produce a Null set. Now the problem is to identify the required set S of minimal LQs from the available LQ cloud so that any learner with given Input knowledge background I can reach to the level R. The condition of minimalism depends upon various factors according to the requirements and preferences of the learner. The metric for minimalism could be duration of learning time, total cost of learning or just the number of LQs in the proposed solution.

Whatever be the condition of minimalism, in order to identify the required set S of minimal LQs, the system has to find an initial set $S_1$ of $k_1$ LQs from the LQC, such that the set union of the objectives of the derived set $S_1$ of LQs minimally covers R, i.e.

$$\bar{U}Oi \supseteq R, (1 \leq i \leq k_1) \qquad (1)$$

There may be zero or more sets of LQs for which the condition 1 mentioned above are true. In the event that no such set of LQs exists which meets condition 1, the LQC under consideration cannot provide any solution for the proposed learner requirement.

After the $k_1$ number of LQs for the initial solution set is made available, we have to do a backward search operation amongst these $k_1$ LQs in the LQ cloud so that we can reach the Input knowledge set (I) of the learner as the pre-requisite(s) of some LQs. However, in this process, if we do not find any such single LQ or multiple LQs, we can conclude that the LQs in set $S_1$ does not provide any sequence of LQs from the existing LQ cloud to form a path for the learner with Input knowledge level I to reach the specified requirement level R.

Once the $k_1$ number of LQs for the initial solution set is made available, we have to look at a sum of individual pre-requisite knowledge for each of these k1 LQs. Let $Q_1'$ represents this, i.e.

$$Q_1' = \bar{U}Pi , (1 \leq i \leq k1) \qquad (2)$$

If the learner's input knowledge set (I) of the learner covers $Q_1'$, then our task is almost over, i.e., we could identify the set of LQs that this learner requires. In other words, the initial set of LQs $S_1$ has been identified as the final and target set S at the end of just a single iteration. However, in a real life situation, more than often we would find that the learner's input knowledge set I cannot cover set $Q_1'$.

In order to find the complete list of LQs from the initial list, backward searching through each of the LQs satisfying 1 is required. This is done as follows:

Considering I as the Input knowledge set of the learner, in order to learn the $k_1$ selected LQs, the revised set of requirements will be,

$$Q_1 = Q_1' - I \qquad\qquad (3)$$

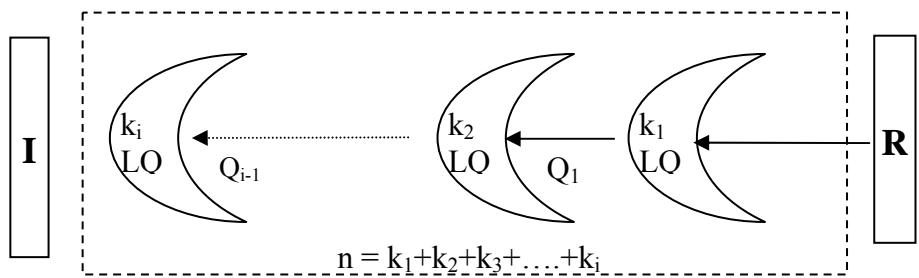

Figure 4: Identifying the required minimal set of LQs

This set of pre-requisites $Q_1$ is beyond the input knowledge of the proposed learner and is available from the LQ cloud under consideration. Thus the system would identify a new set of $k_2$ LQs that minimally covers $Q_1$. The learner, therefore, has to study these $k_2$ additional LQs to build the pre-requisite knowledge level as required to study the set $S_1$.

These new set $S_2$ of $(k_1+k_2)$ LQs in total is thus identified in this second iteration as the revised set of minimal LQs that are to be studied. As the pre-requisite knowledge for the first set of $k_1$ LQs are to be covered by the sum of the Objective sets of $k_2$ additional LQs and I together, the problem reduces down to finding the pre-requisite knowledge for these $k_2$ LQs (see figure 4) and to cross-check if that is covered by the learner's input knowledge I.

At this stage of the processing, the system would take a set union over the individual pre-requisite knowledge for each of these additional $k_2$ LQs to obtain $Q_2'$. The iterative process is continued till a set of $k_g$ additional LQs is identified in the g[th] iteration whose set of pre-requisites, $Q_g$ are covered entirely by I. The set $S_g$ derived up to this stage, is to be marked as the solution set S. If the cardinality of this final solution set S, is n then, n = $(k_1 + k_2 + k_3 + \ldots + k_g)$.

Once the LQs are identified, the sequence through which the learner will proceed must be tracked. For this, we generate pre-requisite directed graph with LQs as node and the edges are formed by the pre-requisite data. Thus if P & Q are two LQs where P is the pre-requisite LQ for Q, we draw an edge from P to Q. If we draw the graph of the LQs as explained above, we may reach to a number of disjoint acyclic digraph with multiple zero-pre-requisite LQs and with multiple finish-LQs.

By the term zero-pre-requisite LQ, we identify those LQs, for which the participant has the Input knowledge, i.e $P_i \subseteq I$ and the finish-LQ represents those LQs for which intersection of R and Oi is non-null and not a pre-requisite of any other LQ in the graph. However, there may be multiple nodes in the graph for which intersection of R and Oi is non-null.

Finally, to derive the sequencing of the LQs we use the topological sort by identifying successively nodes with zero in-degree and removing the edges drawn from it before the second iteration. However, if we reach to nodes with equal sequence number, we can offer those LQs in parallel.

## VI. CONCLUSION

This paper provides the designs that are based on the framework of Learning Technology System Architecture and it is expected that based on this design and its further modification, if any, more efficient and adaptive LMS software product can be developed. This proposed design adapts the advantage of both E-learning and hybrid mode learning and fits into a standard architecture. The LQ approach also ensures person-centric delivery system from a large pool of LQs i.e., LQ cloud.

## REFERENCES

[1.] Xiaofei Liu,Abdulmotaleb Ei Saddik, Nicholas D Georganas, "An Implementable Architecture of an E-Learning System", Electrical and Computer Engineering, 2003. IEEE CCECE 2003. Canadian Conference on 4-7 May 2003, Volume 2,Page(s): 717 - 720

[2.] Alien Corbiere,Chistophe Choquet, "Designer Integration in Training Cycle"URL: http://hal.archives-ouvertes.fr/docs/00/19/03/08/PDF/Alain-Corbiere-2004.pdf

[3.] Learning Technology System Architecture (LTSA) Version 6 URL: ltsc.ieee.org/wg1/files/ltsa_06.doc

[4.] S Ray, N Chaki, R Dasgupta, "Design of an adaptive web-based courseware", IASTED International Conference on Intelligent Systems & Control (ISC 2004), Honulalu, Hawaii, USA, pp266-271,August 23-25,2004

[5.] Sharable Content Object Reference Model (SCORM) 2004 3rd Edition URL :http://www.adlnet.gov/scorm/

[6.] Recommendations for authoring SCORM packages targetting Moodle.
URL : http://docs.moodle.com/en/SCORM

[7.] SCORM content inside of Sakai
URL:http://bugs.sakaiproject.org/confluence/display/SCORM/ Home

[8.] SCORM 2004 3rd Edition Certified Ilias
URL: http://www.adlnet.gov/scorm/certified/

[9.] N Chaki & R Dasgupta, " A learners' Quanta Based Framework for Identification of Requirements and Automated Design of Dynamic Web-based Courseware" 14[th] International Monterey Workshop held at the Monterey Conference Center in Monterey,California, Sept. 10-13,2007

[10.] Carla arnham & Robert Kaleta Learning Technology Center,University of Wisconsin-Milwaukee "Introduction to Hybrid Courses" Volume 8, Number 6: March 20, 2002

[11.] Peter Sands, University of Wisconsin-Milwaukee "Inside Outside, Upside Downside Strategies for Connecting Online and Face-to-Face Instruction in Hybrid Courses " Volume 8, Number 6: March 20, 2002

[12.] S Ray , N Chaki , R Dasgupta " A Learner's Quanta Model based Framework Towards Building Dynamic web-based Courseware " 4[th] International Conference on multimedia and ICTs in Education (mICTE 2006) , Seville, Spain pp 238-242, Nov 22-25, 2006

[13.] Adbel H.,et .al: "Multimedia Integration into a Distance Learning Environment", Proc. Of the #rd International Conference on Multimedia Modeling, Tolouse. (1996)

[14.] Vassileva J, Deters R, " Dynamic Courseware Generation on the WWW" , British Journal onf Educational Technology, Vol.29, No. 1 (1997)

[15.] Taber K.S. , " Development of Student Understanding: A case Study of Stability and Liability in Cognitive Structure" , Research in Science and Technology Education , Vol 13(1) pp. 87-97

[16.] Zhang Goerge Guanaon , "Computer based Knowledge System in the USPTO United States Patent Office , September,2004. URL :http://www.uspto.gov/Patents/UnitedStatesPatent6,795,815.htm

[17.] Munro John, " Facilitating Effective Learning and Teaching ", Proc. Of Technology Colleges Trust Online Conference 2002.

[18.] Cockerton T, Shimell R, "Evaluation of hypermedia Document as a Learning Tool ", Journal of Computer Assisted Learning, Vol 13,No.1 (1997).

[19.] C.Boboila,V.Goerge,S.Boboila, "An Online System for Testing and Evaluation" WSEAS Transaction (Jan 2008)

[20.] S.Encheva,S.Tumin, "Multimedia Factors Facilitating Learning" WSEAS Transaction (Oct 2007)

[21.] S.Jun,S.Kim,W.Lee "Online Pair-Programming for Learning Programming of Novices" WSEAS Transaction (Sept 2007)

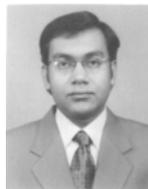

**Souvik Sengupta** had graduated from University of Calcutta, India with Physics honours in 1999. He had done Master of Computer Application from SMU, India in 2002 and Master of Technology from NITTTR under WBUT, India in 2008. He joined as Lecturer in Techno India Group in 2002 and his areas of interests include Web Technology, Object Oriented Technology and Learning Technology.

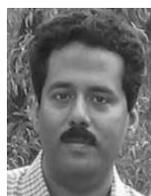

**Nabendu Chaki** is a faculty member in the Department of Computer Science & Engineering, University of Calcutta, Kolkata, India. He received his Ph.D. in 2000 from Jadavpur University, India. His areas of research interests include distributed computing and software engineering. Dr. Chaki has also served as a Research Faculty member in the Ph.D. program in Software Engineering in U.S. Naval Postgraduate School, Monterey, CA during 2001-2002. He is a visiting faculty member for many Universities including the University of Ca'Foscari, Venice, Italy. Dr. Chaki has published close to 50 research papers and a couple of text books. He is a guest editor for the International Journal of Ad Hoc and Ubiquitous Computing (IJAHUC) and has served in the Committees of several International conferences ands workshops.

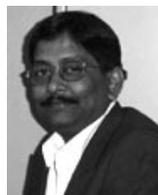

**Ranjan Dasgupta** has joined National Institute of Technical Teachers' Training & Research, Kolkata, India in 1993 as Assistant Professor and presently he is working as Professor & Head, Dept. of CSE. Previously he was with Jadavpur University and also served several Indian IT companies for five years. He received his B. Tech from Dept. of Radio Physics & Electronics, M. Tech & Ph.D from Dept. of Computer Science, University of Calcutta, India. His areas of research interests include software engineering, databases, GIS, distributed computing. Dr. Dasgupta had published around 30 research papers in different International Conferences & Journals. He also served several other Universities & national & international bodies, World Bank Assisted Projects in various capacities.